\documentclass[conference]{IEEEtran}
\IEEEoverridecommandlockouts
\usepackage{cite}
\usepackage{url}
\usepackage{adjustbox}
\usepackage{multirow}
\usepackage{booktabs}
\usepackage{amsmath,amssymb,amsfonts}
\usepackage{algorithmic}
\usepackage{graphicx}
\usepackage{textcomp}
\usepackage{forest}
\usepackage{xcolor}
\def\BibTeX{{\rm B\kern-.05em{\sc i\kern-.025em b}\kern-.08em
    T\kern-.1667em\lower.7ex\hbox{E}\kern-.125emX}}
\begin{document}

\title{Deep scattering network for speech emotion recognition}

\author{\IEEEauthorblockN{Premjeet Singh$^1$, Goutam Saha$^1$, Md Sahidullah$^2$}
\IEEEauthorblockA{$^1$\textit{Dept of Electronics and ECE, Indian Institute of Technology Kharagpur, Kharagpur, India} \\
$^2$\textit{Universit\'{e} de Lorraine, CNRS, Inria, LORIA, F-54000, Nancy, France} \\
premsingh@iitkgp.ac.in, gsaha@ece.iitkgp.ac.in, md.sahidullah@inria.fr}
}

\maketitle

\begin{abstract}
This paper introduces scattering transform for speech emotion recognition (SER). Scattering transform generates feature representations which remain stable to deformations and shifting in time and frequency without much loss of information. In speech, the emotion cues are spread across time and localised in frequency. The time and frequency invariance characteristic of scattering coefficients provides a representation robust against emotion irrelevant variations e.g., different speakers, language, gender etc. while preserving the variations caused by emotion cues. Hence, such a representation captures the emotion information more efficiently from speech. We perform experiments to compare scattering coefficients with standard mel-frequency cepstral coefficients (MFCCs) over different databases. It is observed that frequency scattering performs better than time-domain scattering and MFCCs. We also investigate layer-wise scattering coefficients to analyse the importance of time shift and deformation stable scalogram and modulation spectrum coefficients for SER. We observe that layer-wise coefficients taken independently also perform better than MFCCs. 

\end{abstract}


\section{Introduction}
The introduction of deep learning has caused rapid development in different speech signal processing tasks. One such domain is \emph{speech emotion recognition} (SER) which finds applications in health-care systems, sentiment analysis, and many other human-computer interaction applications \cite{SHAHFAHAD2021, akccay2020speech, el2011survey}. Even after two decades of research, SER is still considered as a challenging task \cite{akccay2020speech}. One of the major challenges is the interpersonal and intrapersonal variability in emotion expression, e.g., different speaking styles, cultural background, language, context, speaker's mood etc \cite{SHAHFAHAD2021, el2011survey}. This causes SER systems to face difficulty in generalization across different emotion speech samples, especially when the train and test data samples belong to different domains~\cite{Parry2019}. This downside motivates the search for a robust feature that can extract emotion-specific information irrespective of different variabilities.

The deep learning based SER approaches generally addresses the issue of variabilities by using deep convolutional neural networks (CNN), which are considered efficient feature extractors in terms of invariances learned against signal shifts \cite{abdelhamid, mao2014}. However, the complexity and \emph{lack of explainability} of deep CNNs come as an added disadvantage to automatic feature extraction ability \cite{lipton2018mythos}. 

To obtain an end-to-end framework, some SER studies use 1D CNNs to learn the relevant features directly from raw speech \cite{trigeorgis}. Such studies observe that deep networks learn representations similar to handcrafted features \cite{Muckenhirn2019}. However, due to inherent instability and requirement of large data samples for training, suitability of deep neural networks becomes uncertain \cite{rolnick2017deep}.

Following similar lines, we propose the use of \emph{scattering transform} for SER.  Scattering transform was introduced in~\cite{ mallat2010recursive, mallat2012group} as a deep convolutional network involving convolution with \emph{predefined kernel} instead of automatic kernel learning. It also introduces stability to both temporal shift and deformations in the feature representation of signals. As emotion cues spread temporally in speech, such characteristics of scattering transform should provide a representation robust to both time-shifting of cues and emotion irrelevant temporal variations. In \cite{anden2014}, authors compute scattering coefficients across the log-frequency domain for frequency transposition invariance to obtain speaker-independent representation for speech recognition. Such characteristics may also help in learning speaker-independent emotion cues.

Scattering transform is used in both 1-D and 2-D data processing. Regarding 1-D signals, authors in \cite{anden2019} introduce joint time-frequency scattering that incorporates multiscale frequency energy distribution over time-invariant representation for various audio classification frameworks. Authors in \cite{ghezaiel:hal-03086433} use two-layer scattering coefficients with CNN layers to obtain a stable descriptor of speaker information from raw speech. In \cite{bruna2013audio}, authors compute different moments of scattering coefficients layers and found that such moments contain enough information for decent voice synthesis quality. Authors in \cite{salamon2015} used scattering transform for urban sound classification and report a marginal performance improvement but with reduced training data. They attribute their finding to better background information learning ability of temporal modulations. Along similar lines, \cite{bauge2013} also applies to scatter coefficients for environmental sound classification. In \cite{lostanlen2021time} a joint time-frequency based scattering representation is used to analyze the timbral similarity between different acoustic instruments. 
Our work is the first to apply scattering transform for SER. We conduct experiments on three different SER datasets using time and frequency scattering coefficients and achieve noticeable improvement over standard mel-frequency cepstral coefficients (MFCCs). The main contributions of this work are summarized below:

\begin{enumerate}
    \item Optimization of scattering transform parameters and analysing its performance for SER,
    \item Analysis of time-domain and frequency-domain scattering coefficients over different databases, and
    \item Layer-wise analysis of  scattering coefficients for SER.
\end{enumerate}

\section{Scattering Transform}
 For 1D signals, the idea behind scattering transform is to obtain a robust feature that remains invariant to the location of information cues (translation invariance) and time warping (diffeomorphism) of information across different instances of the signal.

\subsection{Notion of Lipschitz continuity and information loss in MFCC}
Let the signal $x(t)$ be translated by $c$. Then we have, $x_c(t)=x(t-c)$. Taking its Fourier transform, we obtain, $X_c(\omega)=e^{-j\omega c}X(\omega)$. Taking modulus on both sides, we obtain, $|X_c(\omega)|=|X(\omega)|$. In short time Fourier transform (STFT), translation $c$ is localized to the time frame duration $T$. If $|c|\ll T$, the STFT representation is already invariant to such translation. However, a robust feature representation also requires stability to time deformations appearing in the signal. Let $x(t)$ is warped (deformed) in time by a factor $\tau$, i.e., $x_\tau (t)=x(t-\tau (t))$. A transformation $\phi$ of $x(t)$ is said to be stable to deformation $\tau$ if, $||\phi(x)-\phi(x_\tau)|| \leq C \underset{t}{sup} |\tau ^{'}(t)|||x||$, i.e., the distance between the non-deformed ($\phi(x)$) and deformed ($\phi(x_\tau)$) feature in transformation space $\phi$ should be less than a factor $C$ of maximum amplitude/size of deformation ($\underset{t}{sup} |\tau ^{'}(t)|$). This is also known as Lipschitz continuity condition. Applying this to Fourier transform, we have, $|||X(\omega)|-|X_\tau (\omega)||| \leq C \epsilon ||x||$, where, time deformation $\tau (t)=\epsilon t$. Now the Fourier transform of $x_\tau (t)= x(t-\epsilon t)$ is given as $X_\tau(\omega)= \frac{1}{1-\epsilon} X(\frac{\omega}{1-\epsilon})$. Hence, the frequency components at $\omega$ are shifted to $\frac{1}{1-\omega}$. This effect is more prominent at high frequencies. Hence, modulus of Fourier transform is not Lipschitz continuous. In MFCC, filters with non-linearly varying bandwidths are applied over the STFT. As output of every mel-filter is averaged to get a single coefficient, filter application can be viewed as averaging performed over frequency domain which counters the effect of instability. The mel-filters applied around high frequency regions have higher bandwidth which reduces the effect of deformation instability. However, this averaging in frequency also leads to information loss, especially at higher frequencies. Now, consider $x_t(u)$ to be the signal frame at time $t$. Then $x_t(u)=x(u)\phi(u-t)$ where $\phi$ is a window of length $T$. The application of mel-filters over Fourier transform of $x_t(u)$, can be written in frequency as,

\begin{equation}
Mx(t,\lambda) = \frac{1}{2\pi} \int |x_t(\omega)|^2|\psi_\lambda (\omega)|^2 d \omega
\label{eq1}
\end{equation}

where $x_t(\omega)$ is the frequency response of $x_t(u)$ and $\psi_\lambda(\omega)$ is the mel-filter with support $\lambda$. Since multiplication in frequency becomes convolution in time, Eq. \ref{eq1} becomes, 
\begin{align}
\begin{split}
    Mx(t,\lambda) = \int |x_t * \psi_\lambda(v)|^2dv
    \\
    = \int \left|\int x(u)\phi(u-t)\psi_\lambda(v-u)du\right|^2 dv
\end{split}
\end{align}

If $T$ is much greater than the filter support $\lambda$ in time, $\phi(t)$ is constant over $\lambda$. Hence we can write, $\phi (u-t)\psi_\lambda(v-u)\approx \phi(v-t)\psi_\lambda(v-u)$, which gives,

\begin{align}
\begin{split}
    M_x(t,\lambda)\approx \int \left| \int  x(u)\psi_\lambda (v-u)du \right|^2 |\phi(v-t)|^2dv
    \\
    = |x*\psi_\lambda|^2*|\phi|^2(t)
\end{split}
\label{eq3}
\end{align}

Eq. \ref{eq3} shows that averaging in frequency domain is equivalent to time-domain averaging of $|x*\psi_\lambda|^2$ with window duration $T$ and becomes the basis for scattering transform. Hence, $M_x(t,\lambda)$ remains invariant to time shifts smaller than $T$. In MFCC, $T$ generally equals $20$ms which limits the capturing of long time scale information. If $T$ is increased, loss of high frequency information takes place due to averaging \cite{anden2014}.

\subsection{Layer-wise scattering coefficients}
\label{layerwisesubsection}
To prevent the information loss appearing in MFCC while, at the same time, capturing high frequency information, scattering transform computes second layer coefficients or the modulation spectrogram of signal. This includes computing the frequency response of the time-series of scalogram frequency bins. This is followed by averaging of both scalogram and modulation spectrum coefficients for stability against deformation. Scattering transform uses constant quality factor wavelet filters over complete frequency range to compute scalogram and modulation spectrogram. If $Q$ defines the number of wavelets per octave, the center frequencies of wavelets becomes $\lambda=2^{\frac{k}{Q}}$, where $k$ is an integer and the bandwidth is proportional to $1/Q$. Every wavelet has a support of bandwidth $\lambda/Q$ around center frequency $\lambda$. The same wavelet in time provide a spread of $\frac{2\pi Q}{\lambda}$. To make sure that this spread is less than $T$ (because $T$ is the time window over which time averaging will be computed for stable feature representation) wavelets are only defined for $\lambda \geq \frac{2\pi Q}{T}$. For frequencies less than $\frac{2\pi Q}{T}$  the frequency interval is covered with wavelets of equal bandwidth given by $\frac{2\pi}{T}$, hence covering complete frequency range. The mel-filter operation and filtering due to time window ($T$) in MFCC is replaced with wavelet-based filters in scattering transform.

An issue with using wavelets is that they "commute with translation" \cite{Mallatscatnet}. However, invariance to translation can generally be introduced by averaging. Following this, $\int x(t)*\psi_\lambda(t)dt$ should be translation invariant. But due to symmetric nature of wavelet, $\int x(t) * \psi_\lambda(t)dt=0$. Hence, a modulus non-linear operator is applied to prevent this vanishing of integral, i.e. $\int |x(t)*\psi_\lambda(t)|dt$. The reason behind choosing modulus non-linearity is because modulus operation is contractive and preserves the norm of the vector to which it is applied \cite{anden2019}. The modulus of Fourier transform coefficients are then low-pass filtered with filter $\phi$ to obtain feature representation robust against time-warping induced deformation instability. These features are then invariant to translations smaller than $2^J$, where $J$ is the scale of the low-pass wavelet filter ($\phi$). Therefore, any $m$th order scattering coefficients are given as,
\begin{equation}
    S_{J_m} x(u) = U_m x*\phi_{2^J}(u) = \int U_m x(v)\phi_{2^J}(u-v)dv
\end{equation}

where,
\begin{equation*}
    U_m x = U[\lambda_m]...U[\lambda_2]U[\lambda_1]x = |||x*\psi_{\lambda 1}|*\psi_{\lambda 2}|...|*\psi_{\lambda m}|
\end{equation*}

To obtain invariance to frequency transpositions, wavelet transform is first computed over the log-frequency axis (e.g., $log({\lambda_1}$)) of time-domain scattering coefficients followed by wavelet averaging. This is similar to the DCT operation in MFCC \cite{anden2014}. Such averaging introduces invariance to shifts in frequency defined by the scale of the averaging wavelet. The frequency scattering coefficients are cascaded with time scattering coefficients to generate a feature representation of the signal.

\begin{figure}[t]
    \centering
    \hspace{-0.4cm}\hbox{\includegraphics[scale=0.215]{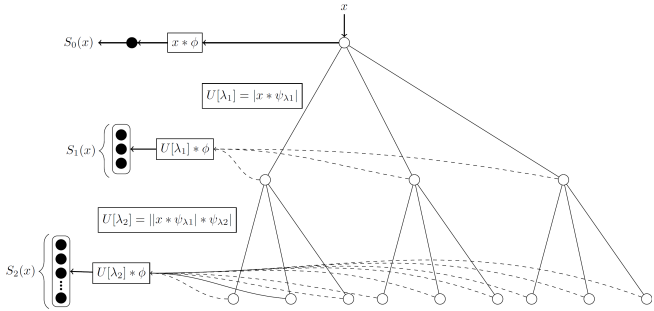}}
    \caption{Graphical representation of scattering transform. Here, $x$ is the input signal and $\phi$ is the low-pass filter used for stability to deformations. $\psi_1$ and $\psi_2$ are wavelet filter banks corresponding to $1$st and $2$nd layer of scattering transform. The figure describes only $2$ layer decomposition of signal $x$.}
    \label{fig:scat_graph}
\end{figure}

\section{Experimental Details}
\subsection{Dataset description}
Three different speech corpora used in the experiments are described in Table \ref{datatable}. The speech samples are downsampled at 16~kHz sampling frequency when required. EmoDB database is a German-language database whereas, both IEMOCAP and RAVDESS contain speech samples in the English language. For IEMOCAP, we use only four emotions (Happy, Angry, Sad, and Neutral) following other works with this dataset~\cite{Parry2019, luo2018investigation}.

\begin{table}[ht]
\centering
\caption{Summary of the speech corpora used in the experiments. (F=Female, M=Male)}
\vspace{0.25cm}
\begin{scriptsize}
\begin{adjustbox}{width=0.5\textwidth,center}
\label{databasetable}
\begin{tabular}{@{}ccccccc@{}}
\toprule
\textbf{Databases} & \textbf{Speakers} & \textbf{Emotions}  \\ \midrule
\begin{tabular}[c]{@{}c@{}}Berlin Emotion Database \\ (EmoDB) \cite{burkhardt2005database}\end{tabular} & \begin{tabular}[c]{@{}c@{}}10\\ (5~F, 5~M)\\ \end{tabular}  & \begin{tabular}[c]{@{}c@{}}7\\ (Anger, Sad, Boredom, Fear,\\ Happy, Disgust and Neutral)\end{tabular} \\ [0.6cm]

\begin{tabular}[c]{@{}c@{}}Ryerson Audio-Visual Database \\ of Emotional Speech and Song \\(RAVDESS)  \cite{livingstone2018ryerson}\end{tabular} & \begin{tabular}[c]{@{}c@{}}24\\ (12~F, 12~M)\end{tabular} & \begin{tabular}[c]{@{}c@{}}8\\ (Calm, Happy, Sad, Angry, Neutral, \\ Fearful, Surprise, and Disgust)\end{tabular}  \\ [0.6cm]

\begin{tabular}[c]{@{}c@{}}Interactive Emotional Dyadic \\ Motion Capture Database \\ (IEMOCAP) \cite{busso2008iemocap} \end{tabular} & \begin{tabular}[c]{@{}c@{}} 10 \\(5~F, 5~M) \end{tabular} & \begin{tabular}[c]{@{}c@{}} 4 \\ (Happy, Angry, Sad and Neutral) \end{tabular}  \\ [0.6cm]

\bottomrule
\end{tabular}
\end{adjustbox}
\label{datatable}
\end{scriptsize}
\end{table}

\subsection{Experimental evaluation \& Methodology}
We first optimize the scattering transform parameters over the EmoDB database. The optimized parameter set is then used for SER evaluation over other databases. We perform only two-level decomposition as higher level coefficients contain a very small fraction of signal energy~\cite{anden2019}. We optimize the number of \emph{wavelets per octave} ($Q$) and \emph{maximal wavelet length} or \emph{averaging scale} ($T$) while using an empirically chosen fixed value of input signal duration ($N$). For frequency scattering, we select the maximal wavelet length as 32. We choose this to obtain invariance over five octaves, found optimum for EmoDB database. Also, following~\cite{anden2014}, for the first layer time-scattering coefficients, we compute only the frequency-domain wavelet decomposition. The required optimum frequency averaging is computed and applied by the classifier itself \cite{anden2014}. We use Morlet wavelet for both time and frequency scattering coefficients computation \cite{anden2014}. For implementation, we use the ScatNet toolkit\footnote{\url{https://github.com/scatnet/scatnet}}.

For classification purposes, we use a radial basis function (RBF) kernel-based \emph{support vector machine (SVM)}, as the feature representations obtained are utterance-level representations. We use the leave-one-speaker-out (LOSO) cross-validation strategy and report average performance obtained over different train-test pairs. For every experiment, we keep speech samples of one speaker in validation, one in test and remaining in training set. We optimize the hyperparameters of the SVM classifier on the validation set.

For comparison, we also compute MFCCs for speech utterances of duration $N$ and evaluate the performance with the same LOSO cross-validation strategy. We choose the MFCC feature because of its characteristic of mimicking human sound perception and its versatility in various speech processing domains. For performance analysis, we use both \emph{accuracy} and \emph{unweighted average recall} (UAR) performance metrics. The UAR is defined as, 
 \begin{equation}
    \mathrm{UAR} = \frac{1}{K} \sum_{i=1}^{K} \frac{A_{ii}}{\sum_{j=1}^{K} A_{ij}}.
\end{equation}

\noindent Here $A$ is the contingency matrix, $A_{ij}$ is the number of samples in class $i$ classified into class $j$ and $K$ is the number of classes. UAR is better suited as a performance metric in class imbalance situations as compared to standard accuracy~\cite{rosenberg2012classifying}.

\begin{figure*}[ht!]
    \centering
    \setlength\abovecaptionskip{-0.5\baselineskip}
    \includegraphics[scale=0.35]{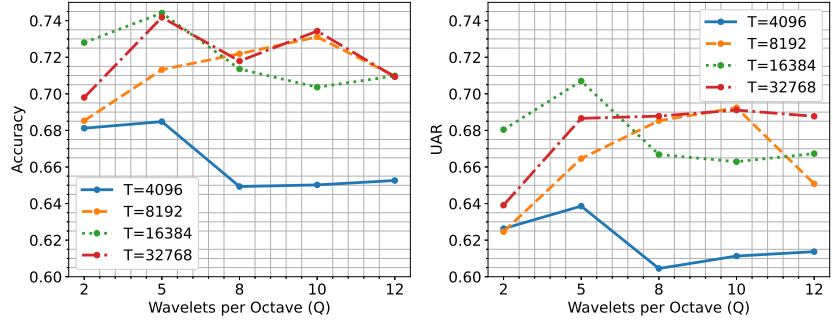}
    \caption{Comparison of various time-domain scattering transform parameters over EmoDB database. $Q$ is number of wavelets per octaves and $T$ is the averaging scale in samples. Improved performance over lower values of $Q$ shows the importance of coarse level frequency-domain information for SER.}
    \label{paramcomp}
\end{figure*}

\begin{figure*}[h!]
    \centering
    \hspace{0.5cm}
    \begin{minipage}{0.315\textwidth}
    \includegraphics[scale=0.35]{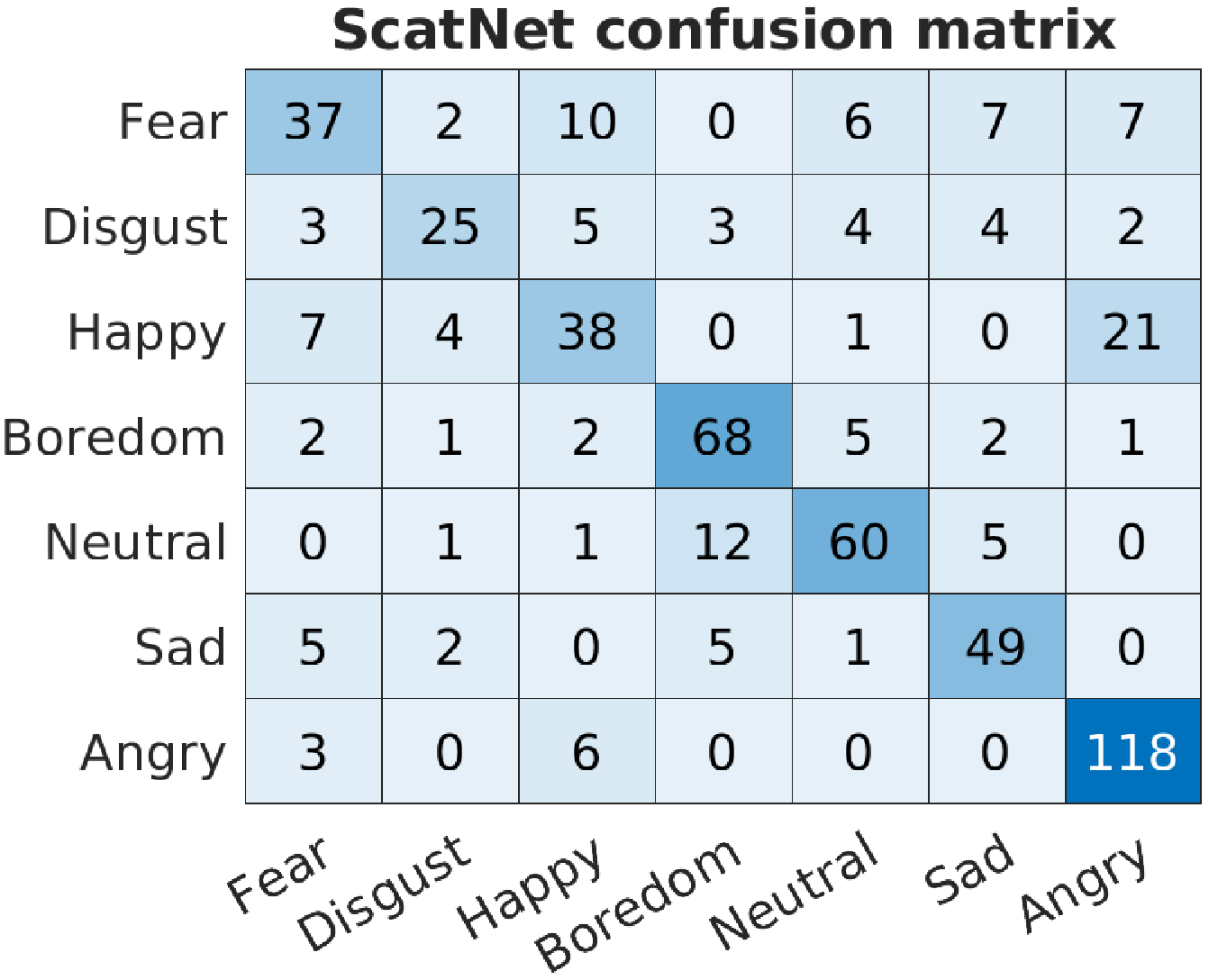}
    \end{minipage}%
    \begin{minipage}{0.315\textwidth}
    \includegraphics[scale=0.35]{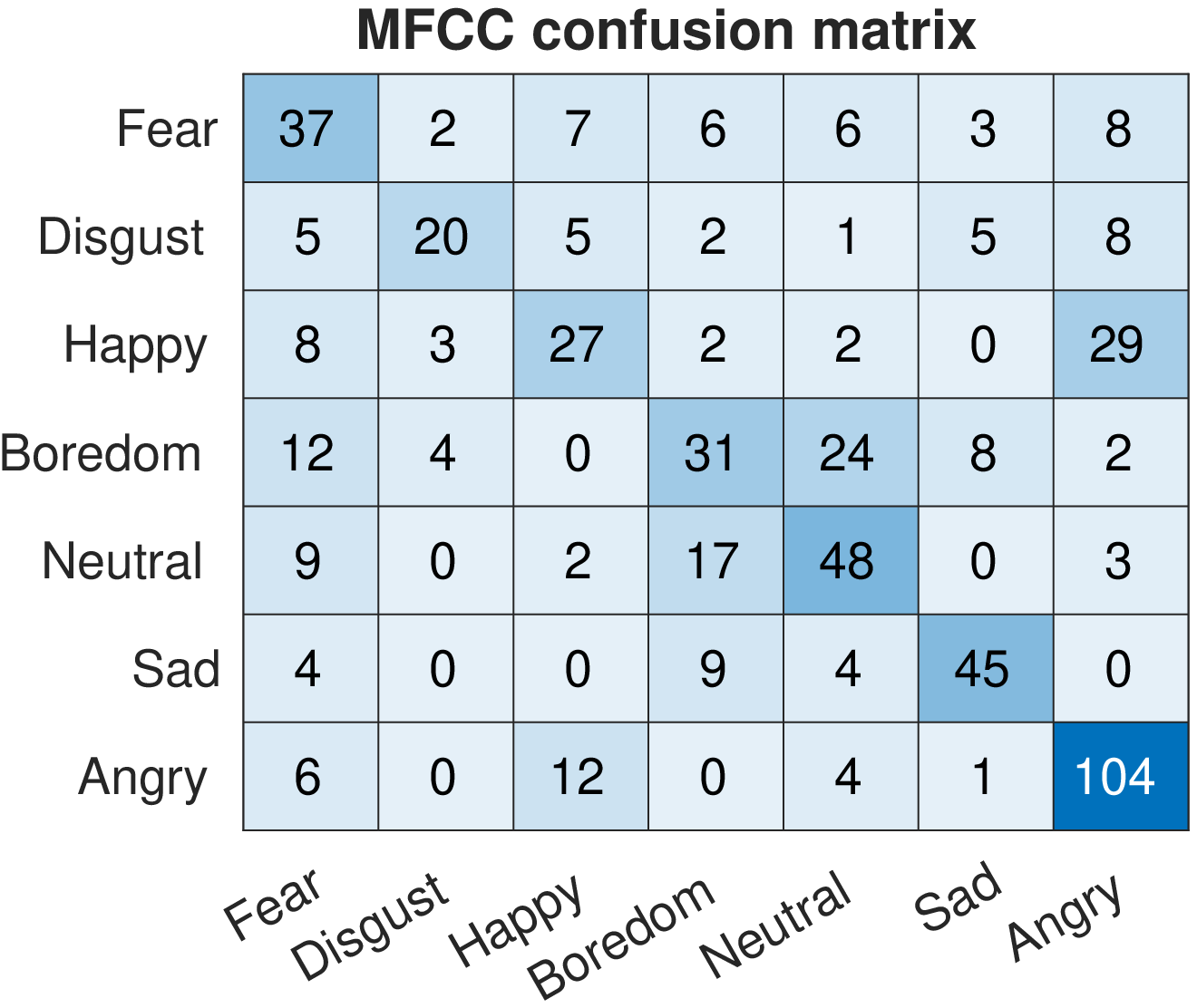}
    \end{minipage}
    \caption{Confusion Matrices of time-domain scattering coefficients and MFCC features over EmoDB database. Although scattering coefficients show higher classification rate for every emotion class as compared to MFCC, confusion across emotion classes of similar arousal characteristics, such as, Happy \& Angry, Neutral \& Boredom and Fear \& Happy can be observed.}
    \label{conf_mats}
\end{figure*}

\section{Results}
\subsection{Parameter optimization over EmoDB}

Figure \ref{paramcomp} shows the results obtained by varying $Q$ and $T$ parameters of scattering transform over EmoDB database. The optimum value of signal duration ($N$) is empirically at $51000$~samples (or $3.18$~seconds at $16$~kHz sampling rate). We start with the lowest value of $T$ to be $4096$~samples, which correspond to $256$~ms at $16$KHz, a duration considered optimum to convey sufficient emotion cues~\cite{dtpmpaper}. $T$ is then increased by a factor of $2$ until its value is below $N$. Interestingly, $T=4096$ shows poor performance, whereas, $T=32768$ shows  better performance over different values of $Q$. In terms of parameter $Q$, we observed a general increase in performance for lower values of wavelets per octave. This indicates the importance of coarse-level frequency domain information for SER. This result is also comparable to the experiments done using constant-Q transform (CQT) for SER~\cite{singh2021non}. We select $T=16384$ and $Q=5$ as the optimum value of parameters for further experiments. Figure \ref{conf_mats} shows the confusion matrices for scattering coefficients and MFCCs over EmoDB database. Scattering coefficients can better classify speech samples of different emotion classes in EmoDB as compared to MFCC.

\subsection{Evaluation on other datasets}

Table \ref{comptable} shows the performance obtained with the optimised scattering parameters (ScatNet) over different databases and its comparison with standard MFCC. We compute 13 MFCCs over window length of $20$~ms, $10$~ms hop and $512$ frequency bins over same utterance duration $N$ ($3.18$ seconds). Mean and standard deviation of MFCCs is computed over all the frames to obtain a vector representation for every utterance. The scattering coefficients are observed to outperform MFCCs over every database. We also compare performance with frequency domain scattering transform (F-ScatNet) with similar optimized parameters ($Q$, $T$ and $N$). The frequency transposition invariance obtained from frequency domain scattering introduces speaker invariance hence improving the performance. However, such improvement is not very prominent in the EmoDB database. One probable reason for such observation could be increased redundancy in frequency scattering coefficients because of small database size and low number of speakers.

\begin{table}[h]
\centering
\renewcommand{\arraystretch}{1.2}
\caption{Performance comparison between frequency scattering (F-ScatNet), time-domain scattering (ScatNet) and MFCC over different SER databases. Given values are in percentages.}
\label{comptable}
\begin{tabular}{|c|c|c|c|c|c|c|c|}
\hline
Database & \multicolumn{2}{c|}{F-ScatNet} & \multicolumn{2}{c|}{ScatNet} & \multicolumn{2}{c|}{MFCC} \\ \cline{2-7} 
\multicolumn{1}{|c|}{} & Acc. & UAR & Acc. &  UAR & Acc. & UAR \\ \hline
EmoDB & 74.59 & 70.27 & 74.40 & 71.30 & 58.39 & 54.03  \\ \hline
RAVDESS & 51.81 & 50.52 & 50.00 & 48.50 & 36.74 & 34.77  \\ \hline
IEMOCAP & 61.55 & 51.00 & 60.41 & 50.40 & 55.54 & 47.19  \\ \hline
\end{tabular}
\end{table}

\subsection{Layer-wise analysis}
To analyse emotion relevance of different layers of scattering transform, we compare SER performance by using the first and second layer coefficients separately in Table \ref{layerwisetable}. The first layer coefficients are averaged scalogram coefficients similar to MFCC. However, the averaging provided by filter $\phi$ introduces time shift invariance which results in better performance than MFCC. The second layer coefficients further perform better than first layer coefficients showing the relevance of modulation spectrum coefficients from SER perspective.

\begin{table}[h]
    \centering
    \renewcommand{\arraystretch}{1.2}
    \caption{SER performance comparison with first and second layer time-domain scattering coefficients taken independently. Given values are in percentages.}
    \label{layerwisetable}
    \begin{tabular}{|c|c|c|c|c|}
    \hline
         Database & \multicolumn{2}{c|}{First layer} & \multicolumn{2}{c|}{Second layer} \\
         \cline{2-5}
         & Accuracy & UAR & Accuracy & UAR \\ \hline
         EmoDB & 62.18 & 57.94 & 68.64 & 61.22 \\ \hline
         RAVDESS & 38.06 & 36.72 & 48.61 & 47.20 \\ \hline
         IEMOCAP & 56.48 & 47.30 & 58.93 & 48.80 \\ \hline
    \end{tabular}
\end{table}

\section{Discussion \& Conclusion}
We observe that the scattering transform coefficients, with optimized parameters, provide feature representations that are better suited for SER. The optimized wavelet averaging scale provides sufficient invariance and stability to irrelevant temporal variations to capture the emotion cues in the time domain. The low value of wavelets per octave generates coarse-level frequency domain information improving SER performance. Invariance to frequency transposition in frequency scattering further reduces the speaker-dependent variations enhancing the system performance. The performed layer-wise analysis shows the importance of time-domain averaging over the typical scalogram/mel-spectrogram coefficients. Results obtained with second layer scattering coefficients indicate the relevance of amplitude modulation of time series of different scalogram frequency bins.

We conclude that deep convolutional scattering coefficients capture more relevant emotion-related information than the standard mel-filterbank based feature. However, the performance with optimized parameter set on EmoDB varies noticeably across databases, indicating the lack of generalization. We also observe that the system faces difficulty in classifying emotions which have similar arousal characteristics (e.g., Happy \& Angry). This hints towards the requirement of further analysis on improving the scattering coefficient based feature representation for SER. Our work is a preliminary study that introduces scattering transform for SER. In future, we will explore other back-end classifiers like multi-layer perceptron, gaussian mixture models etc. with scattering coefficients. To further evaluate the generalisation across samples from different domains, cross-corpus analysis for SER can also be evaluated. 

\bibliographystyle{IEEEtran}
\bibliography{main}

\end{document}